\documentclass[10pt, conference, compsocconf]{IEEEtran}
\usepackage[cmex10]{amsmath}
\usepackage{graphicx}
\usepackage{epsfig}
\usepackage{amssymb,amsmath,array}
\usepackage{url}
\usepackage[ruled,vlined,linesnumbered]{algorithm2e}
\providecommand{\DontPrintSemicolon}{\dontprintsemicolon}

\usepackage{algorithmic}

\DeclareMathOperator*{\argmax}{argmax}

\begin{document}

\title{Feature Selection for Microarray Gene Expression Data\\using
  Simulated Annealing guided by the Multivariate Joint Entropy}

\author{\IEEEauthorblockN{F\'{e}lix Fernando Gonz\'{a}lez-Navarro\IEEEauthorrefmark{1},
Llu\'is A. Belanche-Mu\~noz\IEEEauthorrefmark{2}}
\\
\IEEEauthorblockA{\IEEEauthorrefmark{1}Instituto de Ingenier\'{\i}a\\
Universidad Aut\'{o}noma de Baja California, Mexicali, M\'{e}xico\\
\texttt{fernando.gonzalez@uabc.edu.mx}}
\\
\IEEEauthorblockA{\IEEEauthorrefmark{2}Dept. de Llenguatges i Sistemes Inform\`atics\\
Universitat Polit\`ecnica de Catalunya, Barcelona, Spain\\
\texttt{belanche@lsi.upc.edu}}
}

\maketitle

\begin{abstract}
  In this work a new way to calculate the multivariate joint entropy
  is presented. This measure is the basis for a fast
  information-theoretic based evaluation of gene relevance in a
  Microarray Gene Expression data context. Its low complexity is based
  on the reuse of previous computations to calculate current feature
  relevance.

  The $\mu$-TAFS algorithm --named as such to differentiate it from
  previous TAFS algorithms-- implements a simulated
  annealing technique  specially designed for feature subset selection. 
  The algorithm is applied to the maximization of gene subset
  relevance in several public-domain microarray data sets. The
  experimental results show a notoriously high classification
  performance and low size subsets formed by biologically meaningful
  genes.
\end{abstract}


\section{Introduction}
In cancer diagnosis, classification of the different tumor types is of
great importance. An accurate prediction of different tumor types
provides better treatment and toxicity minimization on
patients. Traditional methods of tackling this situation are primarily
based on morphological characteristics of tumorous tissue
\cite{Chuf05}. These conventional methods are reported to have several
diagnosis limitations. In order to analyze the problem of cancer
classification using gene expression data, more systematic approaches
have been developed \cite{Ying03}.

Pioneering work in cancer classification by gene expression using DNA
microarray showed the possibility to help the diagnosis by means of
Machine Learning or more generally Data Mining methods \cite{Go99},
which are now extensively used for this task \cite{Bo05}. However, in
this setting gene expression data analysis entails a heavy
computational consumption of resources, due to the extreme sparseness
compared to standard data sets in classification tasks \cite{Tan07}.

Typically, a gene expression data set may consist of dozens of
observations but with thousands or even tens of thousands of
genes. Classifying cancer types using this very high ratio between
number of variables and number of observations is a delicate
process. As a result, dimensionality reduction and in particular
\emph{feature subset selection} (FSS) techniques may be very
useful. Finding small subsets of very relevant genes among a huge
quantity could derive in much specific and efficient treatments.

This work addresses the problem of selecting a subset of features by
using the \textsc{TAFS} (Thermodynamic Algorithms for Feature
Selection) family of methods for the FSS problem. Given a suitable
objective function, the algorithm makes uses of an special-purpose
\emph{simulated annealing} (SA) technique to find a good subset of
features that maximizes the objective function. A distinctive
characteristic of TAFS over other search algorithms for FSS is its
probabilistic capability to accept momentarily worse solutions, which
in the end may result in better hypotheses. Despite their powerful
optimization capability, SA-based search algorithms usually lack
execution speed, involving long convergence times. In consequence,
they have been generally excluded as an option in FSS problems, let
alone in highly complex domains such as microarray gene expression
data. A few contributions using the classical SA algorithm for FSS are
found in prostate protein mass spectrometry data \cite{yifen08},
marketing applications \cite{Meiri2006842}, or parameter optimization
in clustering gene expression analysis \cite{Filipone06}.

Our answer to these computational problems is twofold. First, we use a
\emph{filter} objective function for FSS (thus avoiding the
development of a predictive model for every subset
evaluation). Second, the objective function itself is evaluated very
efficiently based in the reutilization of previous computations.

Specifically, a new way to calculate the multivariate joint entropy
for categorical variables is presented that is both exact and very
efficient. This measure is then used by a SA-based \textsc{TAFS}
algorithm to search for small subsets of highly relevant genes in five
public domain microarray data sets. Classification experiments yield
some of the best results reported so far for these data sets and offer
a drastic reduction in subset sizes.

The paper is organized as follows: section II briefly reviews the
Simulated Annealing technique; section III reviews and motivates the
previous Thermodynamic Algorithms for feature subset selection;
section IV develops the information-theoretic measure for feature
relevance and its efficient implementation; section V describes the
data sets and the experimental settings; section VI presents the
results and their interpretation. The paper ends with the conclusions
and directions for future work.

\section{Simulated Annealing}
\label{sec:SimulatedAnnealing} 
Simulated Annealing (SA) is a stochastic technique inspired on
statistical mechanics for finding (near) globally optimal solutions to
large optimization problems. SA is a weak method in that it needs
almost no information about the structure of the search space. The
algorithm works by assuming that some parts of the current solution
belong to a potentially better one, and thus these parts should be
retained by exploring neighbors of the current solution. Assuming the
objective function is to be minimized, then SA would jump from hill to
hill and hence escape or simply avoid sub-optimal solutions.

When a system $S$ (considered as a set of possible states) is in
thermal equilibrium (at a given temperature $T$), the probability that
it is in a certain state $s$, called $P_T(s)$, depends on $T$ and on
the energy $E(s)$ of the state $s$. This probability follows a
Boltzmann distribution:

\begin{eqnarray*}
P_T(s) = \frac{\exp\left(-\frac{E(s)}{kT}\right)}{Z},\ 
with~\ Z = \sum_{s \in S} \exp\left(-\frac{E(s)}{kT}\right)
\end{eqnarray*}

\noindent where $k$ is the Boltzmann constant and $Z$ acts as a
normalization factor. Metropolis and his co-workers developed a
stochastic relaxation method that works by simulating the behavior of
a system at a given temperature $T$ \cite{Metropolis}. Being $s$ the current
state and $s'$ a neighboring state, the probability of making a
transition from $s$ to $s'$ is the ratio $P_T(s \rightarrow s')$
between the probability of being in $s$ and the probability of being
in $s'$:

\begin{eqnarray}
\label{EQN:Prob}
P_T(s \rightarrow s') = \frac{P_T(s')}{P_T(s)} = \exp\left(-\frac{\Delta E}{kT}\right)
\end{eqnarray}

\noindent where we have defined $\Delta E = E(s')-E(s)$. Therefore,
the acceptance or rejection of $s'$ as the new state depends on the
difference of the energies of both states at temperature $T$. If $P_T(s') \geq
P_T(s)$ then the ``move'' is always accepted. It $P_T(s') < P_T(s)$
then it is accepted with probability $P_T(s,s') < 1$ (this situation
corresponds to a transition to a higher-energy state).

Note that this probability depends upon the current temperature $T$
and decreases as $T$ does. In the end, there will be a value of $T$
low enough (the \emph{freezing point}), wherein these transitions will
be very unlikely and the system will be considered frozen. In order to
maximize the probability of finding states of minimal energy at every
value of $T$, \emph{thermal equilibrium} must be reached. To do this,
according to Metropolis, an annealing schedule is designed to prevent
the process from getting stuck at a local minimum. The SA algorithm
introduced in \cite{Kirkpatrick} consists in using the Metropolis idea
at each temperature $T$ for a finite amount of time. In this algorithm
$T$ is first set at a initially high value, spending enough time at it
so to approximate thermal equilibrium. Then a small decrement of $T$
is performed and the process is iterated until the system is
considered frozen.

If the cooling schedule is well designed, the final reached state may
be considered a near-optimal solution. However, the whole process is
inherently slow, mainly because of the thermal equilibrium requirement
at every temperature $T$.

\section{Thermodynamic Algorithms for FSS}
In this section we review TAFS (Thermodynamic Algorithm for Feature
Selection) and eTAFS, two algorithms for FSS that were originally
designed for problems of moderate feature size (up to one hundred)
\cite{TFS08}. If we consider FSS as a search of possible feature
subsets of the full feature set ${\cal X}$, then SA acts as a
combinatorial optimization process \cite{Ree95}. In this sense, TAFS
and eTAFS find a subset of features that optimize the value of a given
objective function $J: {\cal P(X)} \rightarrow \mathbb{R}$. From now
on, we assume that this function is to be maximized and that $J \geq
0$\footnote{This is the case for accuracy, mutual information,
  distances and many other useful measures.}.

To this end, a special-purpose forward/backward mechanism is embedded
into an SA algorithm, taking advantage of its most distinctive
characteristic, the probabilistic acceptance of worse
scenarios over a finite time. This characteristic is enhanced by the
notion of an $\epsilon$-improvement: a feature $\epsilon$-improves a
current solution if it has a higher value of the objective function or
a value not worse than $\epsilon$\%. This mechanism is intended to
account for noise in the evaluation of the objective function (caused
either by the finiteness of the data set or introduced by the chosen
resampling method).

The pseudo-code of TAFS is depicted in \textbf{Algorithm
\ref{CHP4ALG:TAFSI}}. The algorithm consists of two major loops. The
outer loop waits for the inner loop to finish and then updates $T$
according to the chosen cooling schedule. When this loop reaches
$T_{min}$, the algorithm halts. It keeps track of the best solution
found (which is not necessarily the current one).

\begin{algorithm}[!ht]
    \begin{footnotesize}
    \DontPrintSemicolon
    \SetKwInOut{Input}{input}
    \SetKwInOut{Output}{output}
    \Input{
    ${\cal X}:~~$\texttt{Full Feature set} $\{X_{1} \ldots X_{n}\}$\\
    $J():~~$\texttt{Objective Function}\\
    $\alpha():~~$\texttt{Cooling Schedule}\\
    $\epsilon:~~$\texttt{Epsilon}\\
    $T_{0}:~~$\texttt{Initial Temperature}\\
    $T_{min}:~~$\texttt{Final Temperature}\\
    }
    $X_{cur} \leftarrow \emptyset~~$\texttt{Initial current subset}\\
    $J_{cur} \leftarrow 0~~$\texttt{Initial objective function value}\\
    $T \leftarrow T_{0}~~$\texttt{Initial temperature}\\
    \While{$~T>T_{min}$}{
        \Repeat{$\ Y = X_{cur}$}{
            $Y \leftarrow X_{cur}$\\
            $Forward ~~\ (X_{cur},J_{cur})$\\
            $Backward \ (X_{cur},J_{cur})$\\
        }
        $T \leftarrow \alpha(T)$\\
    }
    \end{footnotesize}
    \caption{TAFS algorithm for feature selection}
    \label{CHP4ALG:TAFSI}
\end{algorithm}

\begin{algorithm}[!ht]
    \begin{footnotesize}
    \DontPrintSemicolon
    \SetKwInOut{Input}{input}
    \SetKwInOut{Output}{output}
    \Input{
    $Z,J_Z$
    }
    \Repeat{$not~accept$}{
        $x \leftarrow \displaystyle \argmax_{X_i \in {\cal X} \setminus Z}J(Z \cup \{X_i\})$\\
        \uIf{$>_\epsilon (Z, x, true)$}{
            $accept \leftarrow true$\\
        }
        \Else{
            $\Delta J \leftarrow  J(Z \cup \{x\}) -J(Z)$\\
            $accept \leftarrow rand(0,1)<e^{\frac{\Delta J}{T}}$\\
        }
        \If{$accept$}{$Z \leftarrow Z \cup \{x\}$}
        \If{$J(Z)>J_{cur}$}{$J_Z \leftarrow J(Z)$}
    }
    \end{footnotesize}
    \caption{Procedure Forward ($Z, J_Z$ are modified)}
    \label{CHP4ALG:TAFSFOR}
\end{algorithm}

\begin{algorithm}[!ht]
    \begin{footnotesize}
    \DontPrintSemicolon
    \SetKwInOut{Input}{input}
    \SetKwInOut{Output}{output}
    \Input{
    $Z,J_Z$
    }
    \Repeat{$not~accept$}{
        $x \leftarrow \displaystyle \argmax_ {X_i \in Z}J(Z \setminus \{X_i\})$\\
        \uIf{$>_\epsilon (Z, x, false)$}{
            $accept \leftarrow true$\\
        }
        \Else{
            $\Delta J \leftarrow  J(Z \setminus \{x\}) -J(Z)$\\
            $accept \leftarrow rand(0,1)<e^{\frac{\Delta J}{T}}$\\
        }
        \If{$accept$}{$Z \leftarrow Z \setminus \{x\}$}
        \If{$J(Z)>J_Z$}{$J_Z \leftarrow J(Z)$}
    }
    \end{footnotesize}
    \caption{Procedure Backward ($Z, J_Z$ are modified)}
    \label{CHP4ALG:TAFSBAK}
\end{algorithm}

\begin{algorithm}[!h]
    \begin{footnotesize}
    \DontPrintSemicolon
    \SetKwInOut{Input}{input}
    \SetKwInOut{Output}{output}
    \Input{
    $Z,x,d$
    }
    \Output{boolean}
    \uIf{$d$}{
         $Z' \leftarrow Z \cup \{x\}$\\
    }
    \Else{
        $Z' \leftarrow Z \setminus \{x\}$\\
    }
    $\Delta x \leftarrow J(Z') -J(Z)$\\
    \uIf{$\Delta x > 0$}{
        $return~true$\\
    }
    \Else{
        $return~\frac{-\Delta x}{J(Z)}<\epsilon$\\
    }
    \end{footnotesize}
    \caption{Function $>_\epsilon$}
    \label{CHP4ALG:TAFSEPS}
\end{algorithm}

The inner loop is the core of the algorithm and is composed of two
interleaved procedures: {\em Forward} and {\em Backward}, that iterate
until an equilibrium point is found. These procedures work
independently of each other, but share information about the results
of their respective search in the form of the current solution. Within
them, FSS takes place and the mechanism to escape from local minima
starts working.  These procedures iteratively add or remove features
one at a time in such a way that an $\epsilon$-improvement is accepted
unconditionally, whereas a non $\epsilon$-improvement is accepted
probabilistically. The pseudo-code for Forward and Backward, and
$\epsilon$-improvement is outlined in \textbf{Algorithms
  \ref{CHP4ALG:TAFSFOR}}, \textbf{\ref{CHP4ALG:TAFSBAK}} and
\textbf{\ref{CHP4ALG:TAFSEPS}}. When {\em Forward} and {\em Backward}
finish their respective tasks, TAFS checks if the current solution is
the same as it was prior to their execution. If this is the case, then
we consider that thermal equilibrium has been reached and $T$ is
adjusted, according to the cooling schedule. If it is not, another
loop of Forward and Backward is carried out.

\subsection{eTAFS: an Enhanced TAFS Algorithm}
A modification to Algorithm \ref{CHP4ALG:TAFSI} aimed at speeding up
relaxation time is presented in this section. The algorithm --named
\emph{e}TAFS, see \textbf{Algorithms \ref{CHP4ALG:TAFS2}} and
\textbf{\ref{CHP4ALG:TAFS2BAK}}-- is endowed with a {\em feature
  search window} (of size $l$) in the backward step, as follows. In
{\em forward} steps always the {\em best} feature is added (by looking
all possible additions). In {\em backward} steps this search is
limited to $l$ tries at random (without replacement). The value of $l$
is incremented by one at every thermal-equilibrium point. This
mechanism is an additional source of non-determinism and a bias
towards adding a feature only when it is the best option available. On
the contrary, to remove one, it suffices that its removal
$\epsilon$-improves the current solution. Another direct consequence
is of course a considerable speed-up of the algorithm. Note that the
design of \emph{e}TAFS is such that it grows more and more
deterministic, informed and costly as it converges towards the final
configuration.

\begin{algorithm}[!ht]
    \begin{footnotesize}
    \DontPrintSemicolon
    \SetKwInOut{Input}{input}
    \SetKwInOut{Output}{output}
    \Input{
    ${\cal X}:~~$\texttt{Full Feature set} $\{X_{1} \ldots X_{n}\}$\\
    $J():~~$\texttt{Objective Function}\\
    $\alpha():~~$\texttt{Cooling Schedule}\\
    $\epsilon:~~$\texttt{Epsilon}\\
    $T_{0}~~$\texttt{Initial Temperature}\\
    $T_{min}~~$\texttt{Final Temperature}\\
    }
    $X_{cur} \leftarrow \emptyset~~$\texttt{Initial current subset}\\
    $J_{cur} \leftarrow 0~~$\texttt{Initial objective function value}\\
    $T \leftarrow T_{0}~~$\texttt{Initial temperature}\\
    $l \leftarrow 2~~$\texttt{Window size (for backward steps)}\\
    \While{$~T>T_{min}$}{
        \Repeat{$\ Y = X_{cur}$}{
            $Y \leftarrow X_{cur}$\\
            $Forward ~~\ (X_{cur},J_{cur},l)$\\
            $Backward \ (X_{cur},J_{cur},l)$\\
        }
        $T \leftarrow \alpha(T)$\\
        $l \leftarrow l+1$\\
    }
    \end{footnotesize}
    \caption{\emph{e}TAFS algorithm for feature selection}
    \label{CHP4ALG:TAFS2}
\end{algorithm}

\begin{algorithm}[!h]
    \begin{footnotesize}
    \DontPrintSemicolon
    \SetKwInOut{Input}{input}
    \SetKwInOut{Output}{output}
    \Input{
    $Z,J_Z,l$
    }
    $A \leftarrow \emptyset; A_B \leftarrow \emptyset$\\
    \Repeat{$not~accept$}{
        \For{$i:=1$\bf{ to }$min(l,|Z|)$}{
            Select $x \in Z \setminus A_B$ randomly\\
            \If{$>_\epsilon (Z, x, false)$}{
                $A \leftarrow A \cup \{x\}$
            }
            $A_B \leftarrow A_B \cup \{x\}$\\
        }
        $X_0 \leftarrow \displaystyle \argmax_{X \in A_B} \{J(Z \setminus \{X\})\}$\\
        \uIf{$X_0 \in A $}{
            $accept \leftarrow true$
        }
        \Else{
            $\Delta J \leftarrow  J(Z \setminus \{X_0\}) -J(Z)$\\
            $accept \leftarrow rand(0,1)<e^{\frac{\Delta J}{t}}$\\
        }
        \If{$accept$}{$Z \leftarrow Z \setminus \{X_0\}$}
        \If{$J(Z)>J_Z$}{$J_Z \leftarrow J(Z)$}
   }
   \end{footnotesize}
   \caption{\emph{e}TAFS Backward procedure ($Z, J_Z$ are
     modified). Note that $X_0$ can be efficiently computed while in
     the {\bf for} loop).}
   \label{CHP4ALG:TAFS2BAK}
\end{algorithm}

\section{Information Theoretic Feature Relevance}
\subsection{Entropy definitions}
Entropy, a main concept in information theory \cite{Sha48}, can be
seen as an average of uncertainty in a random variable.  If $X$ is a
discrete random variable with probability mass function $p$, its
entropy is defined by\footnote{All $\log$ are to base 2.}:

\begin{equation}
H(X) = -\sum_{x} p(x)\,\log p(x) = - E_X [\log\,p(X)]
\end{equation}

\noindent being $E[\,]$ the expectation operator. If a variable ($X$)
is known and another one ($Y$) is not, the \emph{conditional entropy}
of $Y$ with respect to $X$: is the mutual entropy with respect to the
corresponding conditional distribution:

\begin{equation}\label{EC}
H(Y|X) = -\sum_{x} \sum_{y} p(x,y)\log\,p(y|x).
\end{equation}

From these two definitions another concept is build, the \emph{mutual
  information} (MI), which can be interpreted as a measure of the
information that a random variable has or explains about another one.

\begin{equation}\label{IM}
I(X;Y) = H(Y) - H(Y|X) = E_{X,Y} [\log\,\frac{p(x,y)}{p(x)p(y)}].
\end{equation}

The computation of the MI can be extended from the bivariate to the multivariate case of a number $n \geq 2$ of variables, against another one:

\begin{eqnarray}
I(X_1,\ldots,X_n;Y) = \sum_{i=1}^n I(X_i;Y|X_1,\ldots,X_{i-1})\nonumber\\
= H(Y) - H(Y|X_1,\ldots,X_n).
\end{eqnarray}

Conditional MI is expressed in the natural way, by conditioning in (\ref{IM}):

\begin{equation}\label{IMC}
I(X;Y|Z) = H(Y|Z) - H(Y|X,Z)
\end{equation}

The MI has been used with success as for feature selection in machine
learning tasks.  Currently there is no agreed-upon definition of the
general multivariate mutual information $I(X_1;\ldots;X_n)$. An
existent proposal is the \emph{interaction information}, described
e.g. in \cite{Jak04} which, for the case of three variables ${X,Y,Z}$,
is defined as $I(X;Y;Z) = I(X;Y|Z) - I(X;Y)$. The extension to the
multivariate case is in terms of the marginal entropies and is given
by:

\begin{equation}
I(X_1; \ldots; X_n) = -\sum_{\tau \subseteq \{X_1, \ldots, X_n\}} (-1)^{n-|\tau|} H(\tau). \nonumber
\end{equation}

This definition is impractical due to its exponential character. In
the next section, the objective function $J$ takes the form of an
information-theoretic index of relevance based on the multivariate
joint entropy, which has already been used elsewhere
\cite{MICAIpaper}. One of the contributions of this paper resides in a
fast implementation of the calculation and its application to
microarray gene expression data.

\subsection{Incremental Multivariate Joint Entropy}
For a random variable $X$, it is known that the joint entropy obeys the following property:

\begin{eqnarray}\label{CHP4eq:JE}
    H(X,Y)\geq H(X)
\end{eqnarray}

This property says that joint entropy is always at least equal to the entropies of the original system: adding a new variable can never reduce the available uncertainty. If we rewrite (\ref{CHP4eq:JE}) as an equation:

\begin{eqnarray}
    H(X,Y) = H(X) + \triangle_X(Y),
\end{eqnarray}

\noindent then $\triangle_X(Y) \geq 0$ represents the \emph{increment}
in entropy due to the addition of the variable $Y$ to the system. In a
feature selection setting, given $Z$ a class variable, $\tau \subset
{\cal X}$ the current subset and $H(\tau)$ its joint entropy, if a new
feature $X_i \in {\cal X} \setminus \tau$ is considered for possible
inclusion in the current subset:

\begin{eqnarray}\label{CHP4eq:ENTROPINC}
    H(Z, \tau \cup \{X_i\})= H(Z, \tau) + \triangle_{Z,\tau}(X_i)
\end{eqnarray}

It turns out that, to obtain the next calculation, it is
computationally far more advantageous to store $H(Z, \tau)$ and
calculate the quantity $\triangle_{Z,\tau}(X_i)$ than to compute the
full joint entropy $H(Z, \tau \cup \{X_i\})$ directly. In order to
obtain this value, and incremental procedure to calculate multivariate
joint entropy has been developed, as described in the sequel.

\begin{table*}[!htb]
\centering
\begin{scriptsize}
    \begin{tabular}{c c c c c}
        \begin{tabular}{c c c}
            $X_1$&$P(X_1)$&$-P(X_1)\,\log P(X_1)$\\
            \hline
            \hline
            0 & 0.538 &  0.481\\
            1 & 0.462 &  0.515\\
            \cline{3-3}
            &$H(X_1)=$&0.996\\
            \hline
            \hline
        \end{tabular}&&&&
        \begin{tabular}{c c c c}
          $X_1$&$X_2$&$P(X_1,X_2)$&$-P(X_1,X_2)\,\log P(X_1,X_2)$\\
            \hline
            \hline
            0 & 0 & 0.231 &  0.488\\
            0 & 1 & 0.308 &  0.523\\
            \cline{4-4}
              &   & $H(X_1,X_2)=$&1.011\\
            \hline
            1 & 0 & 0.154 &  0.415\\
            1 & 1 & 0.308 &  0.523\\
            \cline{4-4}
              &   & $H(X_1,X_2)=$&0.939\\
            \hline
            \hline
        \end{tabular}
    \end{tabular}
    \caption {\emph{Marginal Entropy Scheme} (MES) tables for one
      variable (left) and the addition of a second variable
      (right). $P(\cdot)$ is the probability mass function,
      obtained from the data (all entropies are in bits).}
\label{CHP4Tab:MES1}
\end{scriptsize}
\end{table*}

The incremental multivariate joint entropy (\ref{CHP4eq:ENTROPINC})
must be computed at every evaluation step involving a possible
candidate feature $X_i$ to be included in the current subset
$\tau$. Throughout the process, $\tau$ is associated with its current
\emph{Marginal Entropy Scheme} (MES), a table storing the unique
values contained in the data set for its forming features and its
corresponding entropy value. An example of a MES table for two binary
variables $\{X_1,X_2\}$ is shown in Table \ref{CHP4Tab:MES1}.

At the initial step ($\tau=\emptyset$) the MES table for the addition
of $X_1$ to $\emptyset$ is indicated in the left part of Table
\ref{CHP4Tab:MES1}. The two unique values and their entropies
$H(X_1=0)=0.481$ and $H(X_1=1)=0.515$ are calculated. Let us suppose
that a feature $X_2$ is to be evaluated w.r.t the current subset
$\tau=\{X_1\}$. The MES table with its unique forming patterns are
indicated in the right part of Table \ref{CHP4Tab:MES1}. We can see
that by introducing $X_2$ to the current subset $\tau$, four
\emph{partitions} are generated for each unique value of $X_1$:
$\{00,01,10,11\}$. In the particular case of $X_1=0$, a change in its
entropy contribution is produced by the action of $X_2$ by splitting
it into two entropy values: $H(X_1=0,X_2=0)=0.488$ and
$H(X_1=0,X_2=1)=0.523$, for a total entropy of
$H(X_1=0,X_2)=1.011$. The increment in entropy $\triangle_{\tau}$ is
obtained as the difference between the current MES (considering the
addition of $X_2$) and the previous scheme (without it) --see Table
\ref{CHP4Tab:MES2}.

\begin{table}[!h]
\centering
\begin{scriptsize}
    \begin{tabular}{c@{}|c@{}|c@{}|c@{}}
        $\triangle_\tau$&$H(X_1,X_2)$&$-P(X_1)\,\log P(X_1)$&difference\\
        \hline
        \hline
        $\triangle_\tau(X_1=0)$&1.011&0.481&0.531\\
        $\triangle_\tau(X_1=1)$&0.939&0.515&0.424\\
        \cline{4-4}
        &&$\triangle_\tau$&0.954\\
        \hline
        \hline
    \end{tabular}
\caption {$\triangle_\tau$ computations from the \emph{Marginal Entropy Scheme} --see Table \ref{CHP4Tab:MES1}.}
\label{CHP4Tab:MES2}
\end{scriptsize}
\end{table}

Finally, this last value is applied to eq. (\ref{CHP4eq:ENTROPINC}) to
obtain the joint entropy
$H(X_1,X_2)=H(X_1)+\triangle_\tau(X_2)=0.996+0.954=1.950$. The
listings in \textbf{Algorithms \ref{CHP4ALG:INCENT}} and
\textbf{\ref{CHP4ALG:INCENT-MES}} show the pseudo-code to compute the
procedure explained above. The notation $D|\tau$ stands for the
restriction of the dataset $D$ to the features in $\tau$.

\begin{algorithm}[!h]
    \begin{footnotesize}
        \DontPrintSemicolon
        \SetKwInOut{Input}{input}
        \SetKwInOut{Output}{output}
        \Input{
        $\tau$: Current subset;\\
        $X_i$: feature to be added;\\
        $H_\tau:$ Current subset joint entropy;\\
        $E_\tau:$ Marginal entropies scheme of $H_\tau$;\\
        $D:$ Data set;}
        \Output{$\tau,~H_\tau,~E_\tau$}
        \uIf{$|\tau|=0$}{
            $\tau \leftarrow \{X_i\}$\\
            $D \leftarrow$ Sort$(D)$\\
            $H_\tau \leftarrow$ Joint Entropy of $D$\\
            $E_\tau \leftarrow MarginalEntropyScheme(D|\tau)$\\
            }
        \uElse{
            $\tau^+ \leftarrow \tau \cup \{X_i\}$\\
            $Sort(D|\tau^+)$\\
            $E_{\tau^+} \leftarrow MarginalEntropyScheme(D|\tau^+)$\\
            $E_{\tau^-} \leftarrow \displaystyle \sum_{j} E_{\tau}^j~~$//$j$ \emph{runs through the values of} $\tau$\\
            %//$E_{\tau^+}^j$ is the MES of the $j$-th partition of $D|\tau^+$\\
            $\triangle_\tau \leftarrow \displaystyle\sum_{i} {E_{\tau^+}^i-E_{\tau^-}^i}~~$\\
            $\tau \leftarrow \tau_+$\\
            $H_\tau \leftarrow H_\tau + \triangle_\tau$\\
            $E_\tau \leftarrow E_{\tau^+}~~$//\emph{ new MES}
        }
    \end{footnotesize}
    \caption{Incremental Multivariate Joint Entropy}
    \label{CHP4ALG:INCENT}
\end{algorithm}

Initial entropy is evaluated in lines 2-5. This first step calculates
starting joint entropy as well as its first MES (lines 4-5), which
will be taken as input to the next computation. Note that these two
lines can be efficiently implemented as one function and using only
one loop-cycle, with complexity $\theta(|D|)$, where $|D|$ is the
number of training instances.

In the \textbf{else} part of the \textbf{if} clause, the MES is
calculated with the addition of $X_i$ to the current subset $\tau$
(named $E_{\tau^+}$). Taking into account that previous MES inherits
the ordering sequence derived from a previous stage (because of lines
5 and 9), entropies generated by changes in the MES given by $\tau
\cup \{X_i\}$ are summed ($E_{\tau^-}$) in groups (line 11) by the
newly formed patterns, rendering a one-to-one correspondence between
previous MES and current MES.

\begin{algorithm}[!h]
    \begin{footnotesize}
        \DontPrintSemicolon
        \SetKwInOut{Input}{input}
        \SetKwInOut{Output}{output}
        \Input{
        $D$   : Data set;\\}
%        $\tau$: Current subset;}
        \Output{$E$}
        \ForEach{unique value $v$ in $D$}{
        $\Upsilon[v] \leftarrow$ fraction of instances in $D$ with value $v$
        }
        $E \leftarrow H(\Upsilon)~~$//\emph{calculate entropy of this distribution}
        
    \end{footnotesize}
    \caption{\emph{MarginalEntropyScheme} Function}
    \label{CHP4ALG:INCENT-MES}
\end{algorithm}

Thus, the entropy contribution $\triangle_{\tau}(X_i)$, showing the
effect of adding $X_i$ to $\tau$, is computed by the difference in
both MESs (line 11), being finally added to the current entropy
$H_\tau$ (line 13). The implementation of lines 10-11 follows the same
consideration as lines 4-5, and hence complexity is in the same order.

The incremental multivariate joint entropy is used to obtain an
\emph{index of relevance} (acting as the objective function) of a
feature $X_i \in {\cal X}$ to a class $Z$ with respect to a subset
$\tau \subset {\cal X} \setminus X_i$ and is defined by:

\begin{eqnarray}
J(X_i;Z|\tau) = \frac{H(Z)+H(\tau,X_i)-H(\tau,Z,X_i)}{H(Z)}
\label{CHP4eq:RIM}
\end{eqnarray}

Note the denominator acts as a normalization factor, such that $J \in
[0,1]$, with $J=1$ corresponding to the highest relevance. The reward
of using this objective function by a TAFS-like algorithm consists in
the possibility of testing it in highly complex domains such as
microarray data sets. We name the combination of $e$TAFS and the
objective function in eq. (\ref{CHP4eq:RIM}) as the $\mu$-TAFS
algorithm.

\section{Experimental Work}
\label{sec:ExperimentalWork}

To compute the necessary entropies described in previous section, a
discretization process is needed. This change of representation does
not often result in a significant loss of accuracy (sometimes
significantly improves it \cite{Man05}, \cite{Pot04}); it also offers
reductions in learning time \cite{Cat91}. In this work, the CAIM
algorithm was selected for two reasons: it is designed to work with
supervised data, and does not require the user to define a specific
number of intervals \cite{Luk04}.

\subsection{Data sets}
Five public-domain microarray gene expression data sets are used to
test and validate the approach proposed in this work: \emph{Colon
  Tumor}: 62 observations of colon tissue, of which 40 are tumorous
and 22 normal, 2,000 genes \cite{Al99}. \emph{Leukemia}: 72 bone
marrow observations and 7,129 probes: 6,817 human genes and 312
control genes \cite{Go99}. The goal is to tell acute myeloid leukemia
(AML) from acute lymphoblastic leukemia (ALL). \emph{Lung Cancer}:
distinction between malignant pleural mesothelioma and adenocarcinoma
of lung \cite{Gor02}; 181 observations with 12,533
genes. \emph{Prostate Cancer}: used in \cite{Sig02} to analyze
differences in pathological features of prostate cancer and to
identify genes that might anticipate its clinical behavior; 136
observations and 12,600 genes. \emph{Breast Cancer}: 97 patients with
primary invasive breast carcinoma; 12,600 genes analyzed \cite{Ver02}.

\subsection{Settings}

Provided that the core nature of the $\mu$-TAFS algorithm, and even
many other algorithms --e.g. Genetic Algorithms, Neural Networks--,
resides in their stochasticity, several runs have to be performed, in
order to better asses the average behaviour of the methods.

The experimental design to test $\mu$-TAFS algorithm measures
performance by carrying out $m=100$ different independent runs. In
each run, $\mu$-TAFS is executed on the corresponding dataset and
returns the set of all those feature subsets reaching the best found
performance function (maximum relevance, in this case). To overcome
the existence of many solutions, the subset that offers the lowest
mutual information (MI) among its elements --i.e. the less
redundancy-- is taken as the subset delivered in this run. After
completing the $m$ execution runs, the obtained subsets can be ordered
from minimum to maximum MI value.

The $\mu$-TAFS parameters are as follows: $\epsilon=0.01, T_0=0.1$ and
$T_{min}=0.0001$. These settings were chosen after preliminary
fine-tuning and are kept constant for all the problems
\cite{TFS08}. The cooling function was chosen to be geometric
$\alpha(t)=0.9\,t$, following recommendations in the literature \cite{Ree95}.

\begin{table}[!ht]
    \begin{center}
    \begin{tabular}{l r r c}
        Data set       &  Time &    Jeval   & size \\
        \hline
        \hline
        Colon Tumor    &  6.41 &  503,901&6.93 $\pm$ 0.06\\
        Leukemia       &  6.51 &  506,489&3.36 $\pm$ 0.06\\
        Lung Cancer    &  7.45 &  560,972&2.58 $\pm$ 0.04\\
        Prostate Cancer& 98.74 & 7,119,800&9.85 $\pm$ 0.05\\
        Breast Cancer  &136.93 &10,943,628&9.62 $\pm$ 0.03\\
        \hline
    \end{tabular}
    \end{center}
    \label{Tab:PERF}
    \caption{$\mu$-TAFS running performance. \emph{Time} indicates the average running time (in minutes) over the 100 executions; \emph{Jeval} is the average number of evaluations of $J$; \emph{\emph{size}} the average size of the final solutions and its standard error.}
\end{table}

\begin{figure}[!h]
\centering
\begin{tabular}{@{~}c@{~}c@{~}}
    \epsfig{file=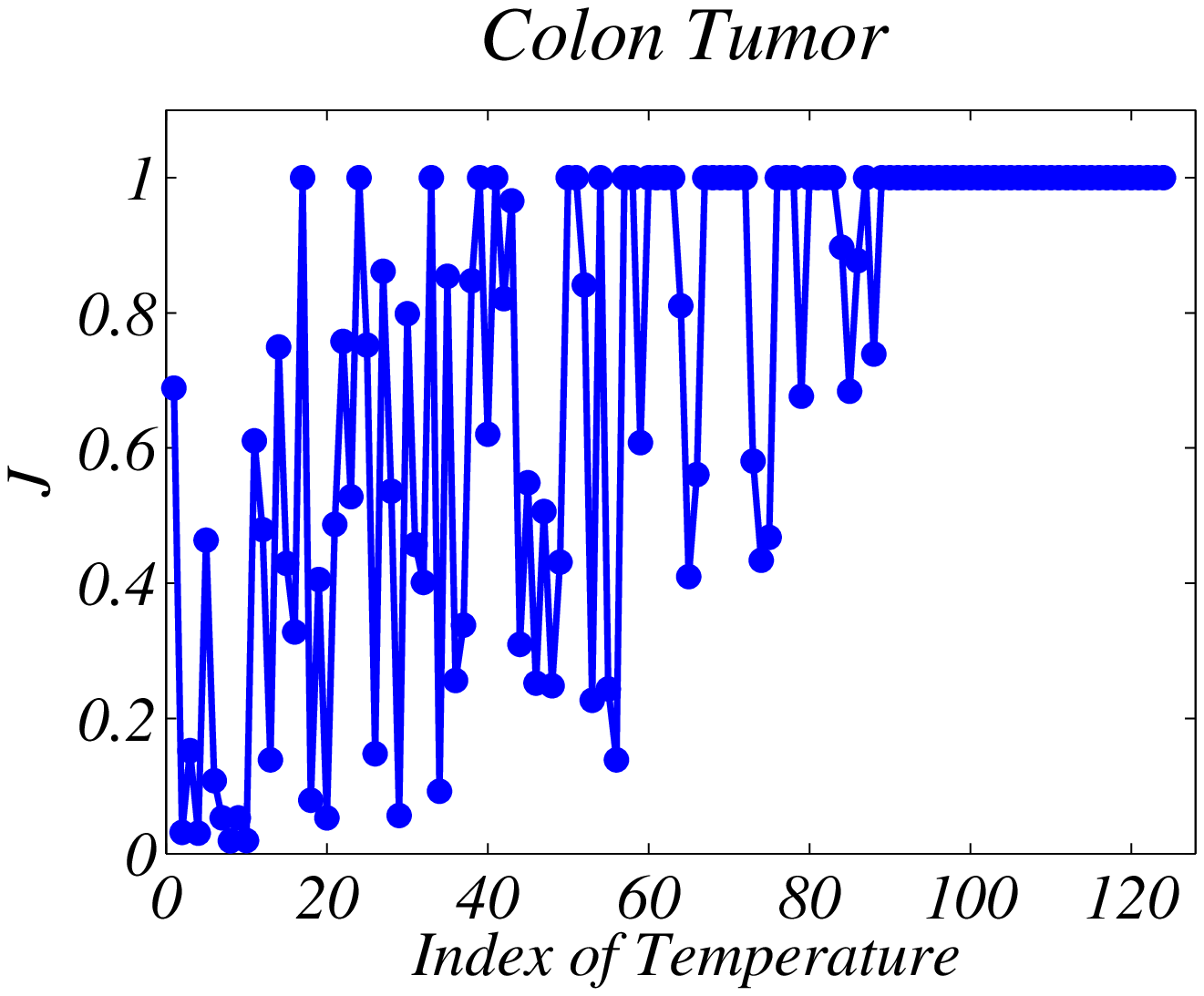,height=3.5cm,bbllx=96,bblly=238,bburx=480,bbury=553}&\epsfig{file=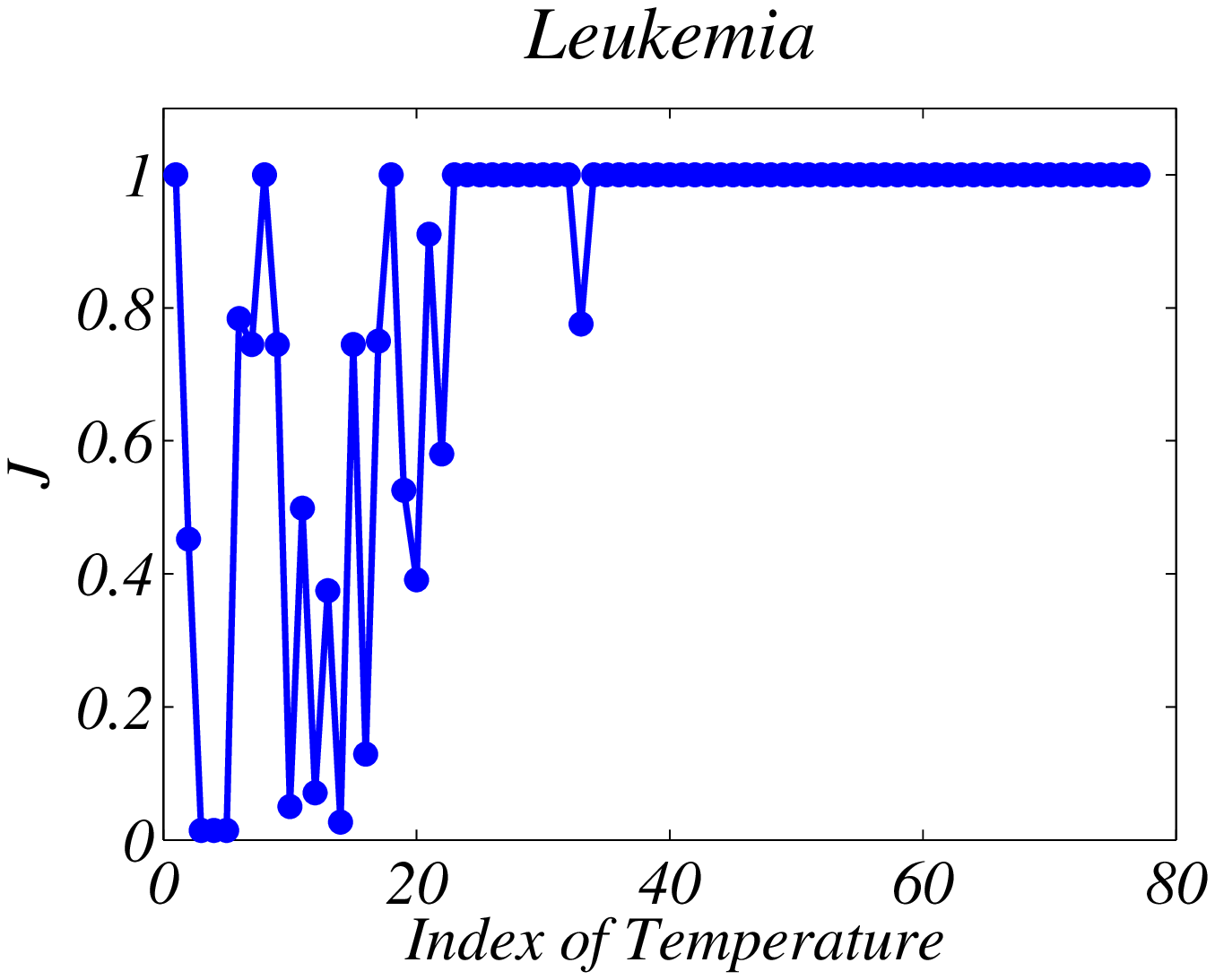,height=3.5cm,bbllx=96,bblly=238,bburx=480,bbury=553}\\
    \epsfig{file=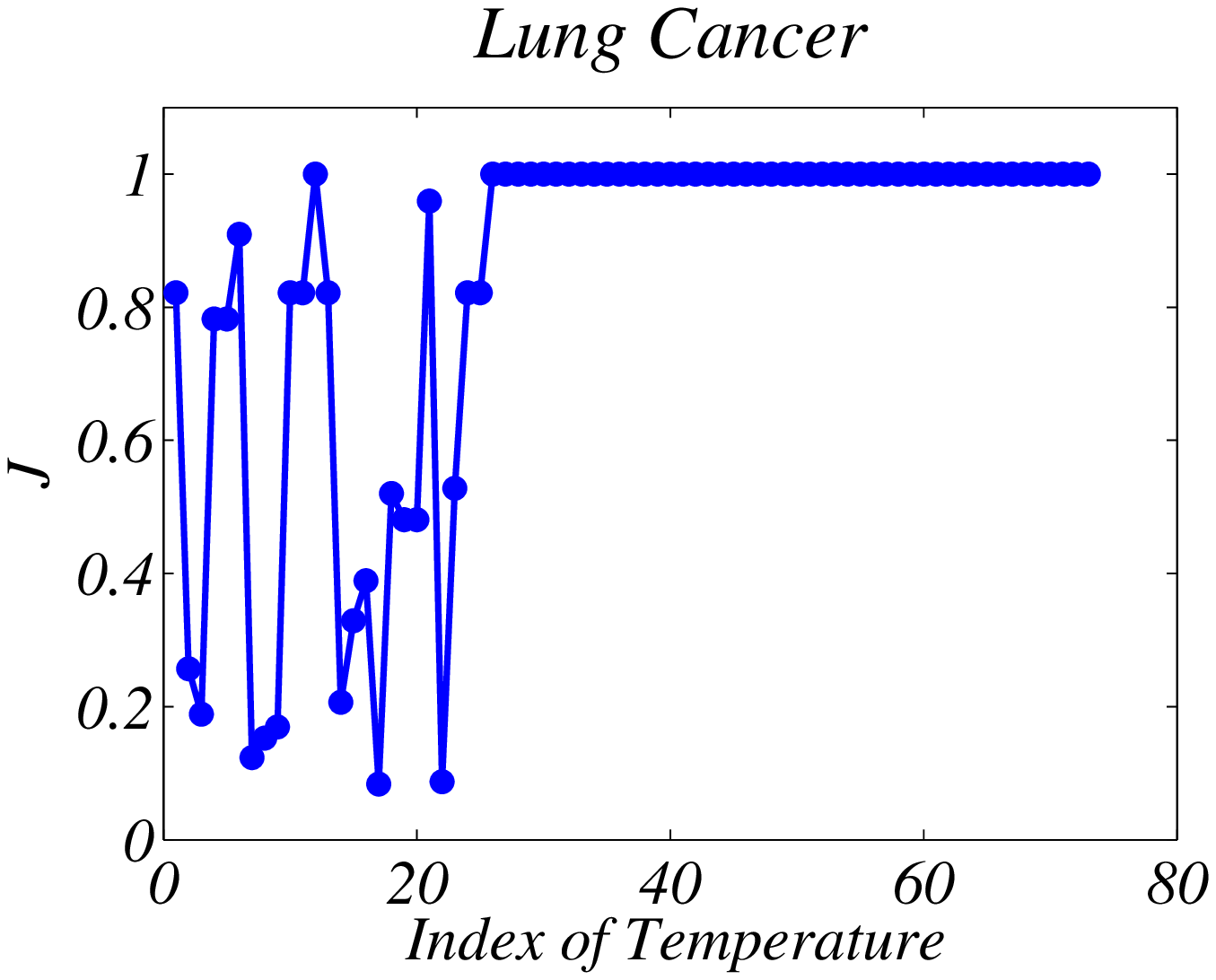,height=3.5cm,bbllx=96,bblly=238,bburx=480,bbury=553}&\epsfig{file=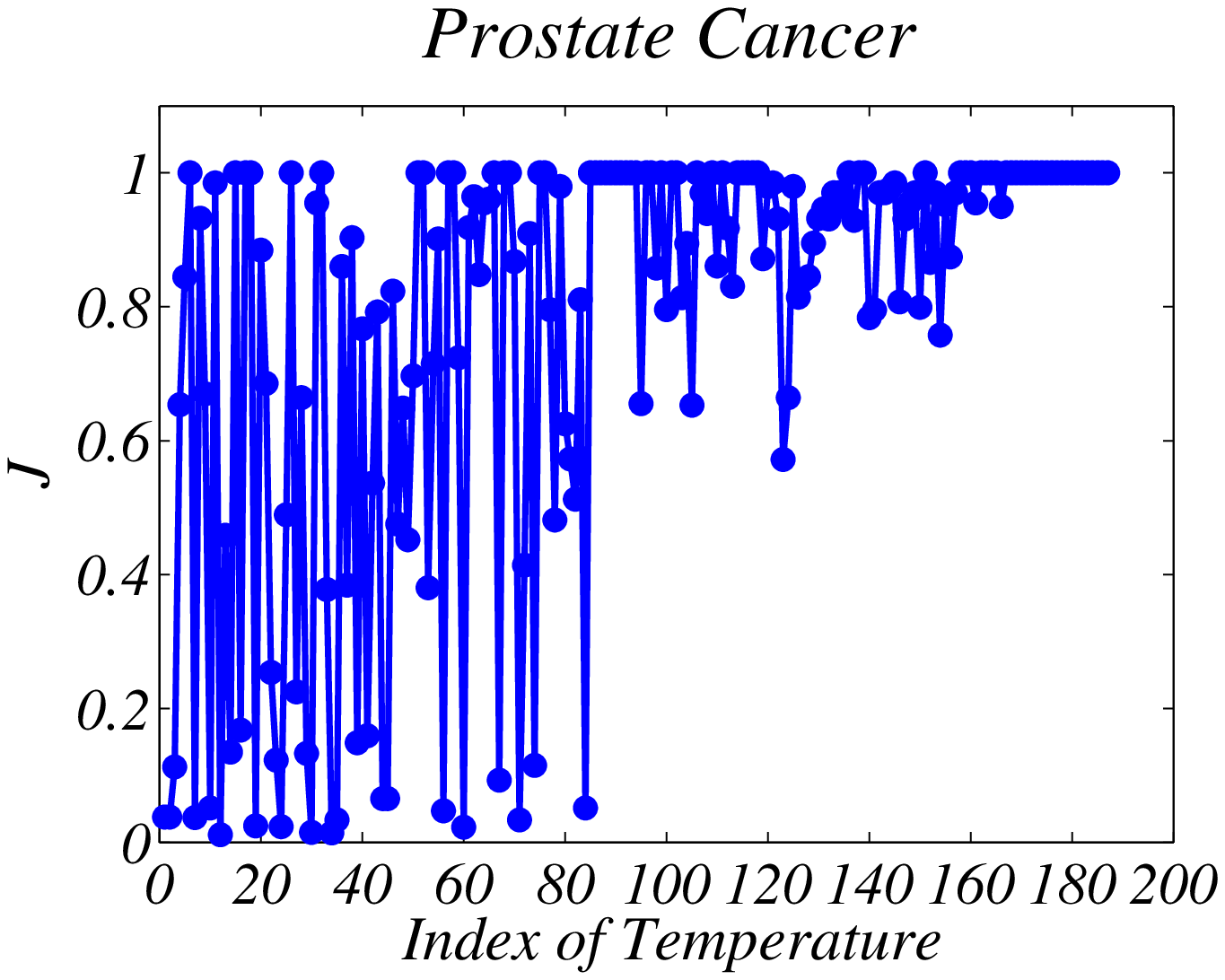,height=3.5cm,bbllx=96,bblly=238,bburx=480,bbury=553}\\    \multicolumn{2}{c}{\epsfig{file=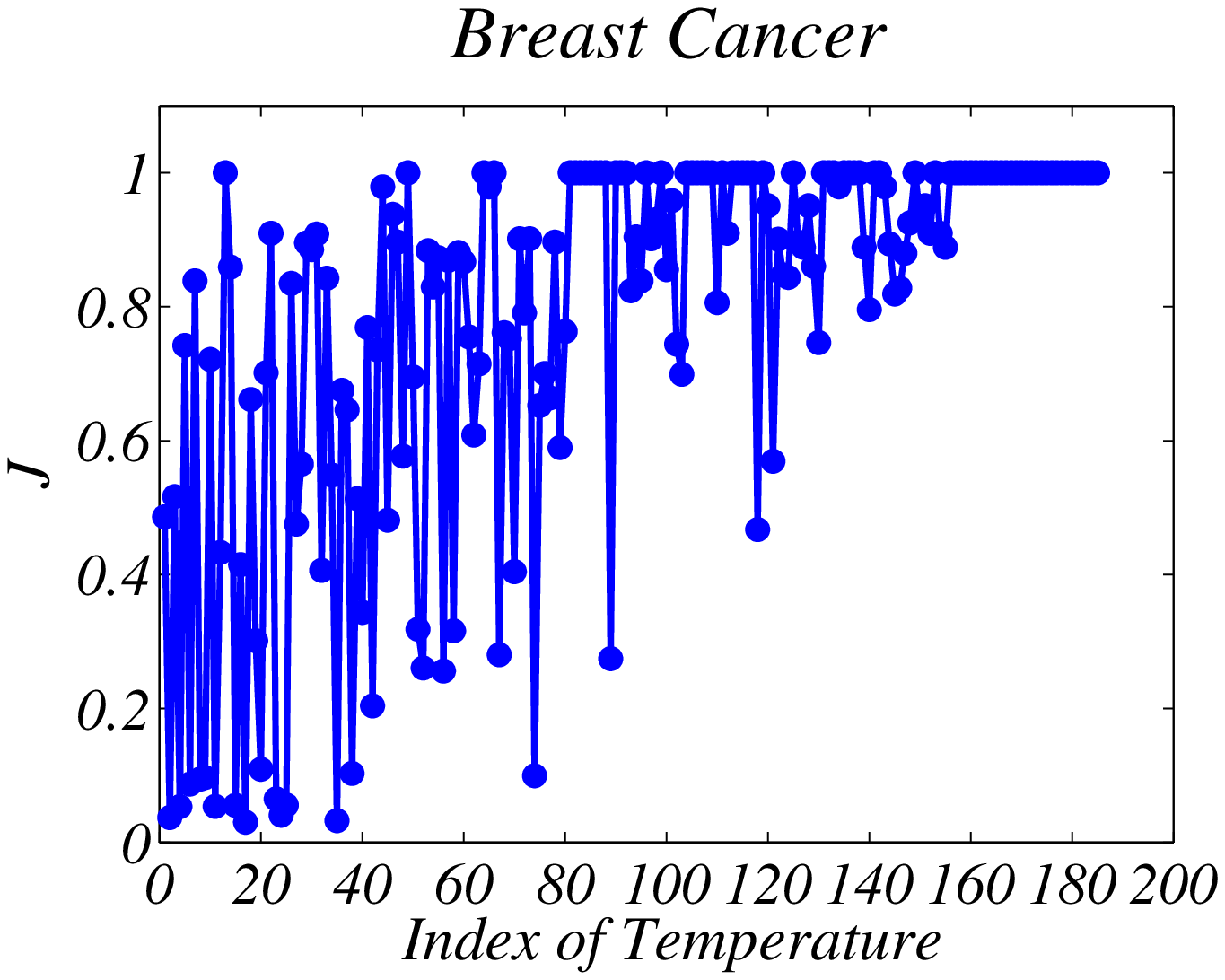,height=3.5cm,bbllx=96,bblly=238,bburx=480,bbury=553}}\\
\end{tabular}
    \caption{$\mu$-TAFS search processes. The x-axis is the iteration counter for the outer loop of the algorithm.}
    \label{FIG:TFS-process}
\end{figure}

\section{Experimental results}

\subsection{$\mu$-TAFS performance results}
The evolution of $\mu$-TAFS from a high temperature state to a frozen
point is depicted in Fig. \ref{FIG:TFS-process}. Highly unstable
--i.e. high temperature condition-- readings are observed at the
initial stages in each of the datasets. As soon as the algorithm
becomes more relaxed due to eq. (\ref{EQN:Prob}), worse solutions are
avoided. The frozen condition is observed at the final stages of each
execution, where $J$ values consecutively reach the maximum possible
value ($J=1$) in all cases.

The running performance of $\mu$-TAFS is summarized in Table
III. The results show that $\mu$-TAFS yields subsets of
considerably low size and also low variability. Notorious readings
correspond to \emph{Leukemia} and \emph{Lung Cancer}. It is
conjectured that such sizes respond to the nature of the proposed
information theoretic model on discretized data sets, in the sense
that only a few genes significantly contribute to increase the index
of relevance given by eq. (\ref{CHP4eq:RIM}). On the one hand, working
with continuous features, the index would tend to vary smoothly
--i.e. generating small increments; as a consequence, more features
are added-deleted. On the other hand, discrete features variations are
\emph{normalized} by their discretization scheme, so small increments
in the real-value are merged into a single discrete value. Therefore,
mostly significant increments are truly reflected in its
addition-deletion from the current subset.

Computational demands when processing the smaller data sets
(as for \emph{Colon tumor}, with 2,000 features) are kept by $\mu$-TAFS
considerable low (5 to 10 minutes). The two more complex
problems, \emph{Prostate} and \emph{Breast Cancer} require
approximately 1.5 and 2 hours of total processing time. Unfortunately,
there is scarcely any reporting on time consumption in recent
scientific literature that would enable us to establish a reasonable
comparison.

\subsection{$\mu$-TAFS accuracy results}

Eight classifiers were evaluated by means of 10 times 10-fold Cross
Validation (10x10 CV), a resampling method designed to handle
small-sized data sets. The chosen classifiers are: the
nearest-neighbor technique with Euclidean metric (kNN) and parameter
\emph{k} (number of neighbors running from 1 to 15), the
\emph{Na\"{\i}ve Bayes classifier} (NB), a {\em Linear and Quadratic
  Discriminant classifier} (LDC), \emph{Logistic Regression} (LR), the
{\em Support Vector Machine with linear and quadratic kernel} (lSVM
and rSVM) and parameter \emph{C-regularization constant} (with
$C=2^k$, $k$ running from $-7$ to $7$) and the {\em Support Vector
  Machine with radial basis function kernel} (rSVM) and parameter
\emph{C} and \emph{$\gamma$-smoothing in the kernel function} (with
$\gamma=2^k$, $k$ running from $-7$ to $7$)\footnote{For the
  experiments, we use a \textsc{MATLAB} implementation; specifically,
  for the SVMs we use the MATLAB interface to \textsc{LIBSVM}
  \cite{libsvm}. All tests are run on on a regular x86
  workstation.}. The non parametric Wilcoxon signed-rank
test\footnote{The Wilcoxon signed-rank test is a non-parametric
  statistical hypothesis test for the analysis of two related samples,
  or repeated measurements on a single sample. It can be used as an
  alternative to the paired Student's t-test when the population
  cannot be assumed to be normally distributed.  It should therefore
  be used whenever the distributional assumptions that underlie the
  t-test cannot be satisfied.} is used for the (null) hypothesis that
the median of the differences between the errors of the winner
classifiers per data set and another classifier's error is zero. The
non-parametric Wilcoxon signed-rank test will be used for the (null)
hypothesis that the median of differences between classifiers
accuracies are zero, at the 95\% level of significance.

% \ref{Tab:PERF}

\begin{table}[!ht]
  \centering
    \begin{scriptsize}
    \begin{tabular}{l c c c c}
         Data set       &   Classifier & 10x10 CV& size \\
        \hline
        \hline
        Colon Tumor     & lSVM $(C=2^1)$          & 89.19$\pm$0.38&5\\
        Leukemia        & lSVM $(C=2^{-7})$       & 99.62$\pm$0.27&3\\
        Lung Cancer     & LR                    & 99.89$\pm$0.07&4\\
        Prostate Cancer & NN (6)                & 95.66$\pm$0.21&7\\
        Breast Cancer   & rSVM $(C=2^{3},\gamma=2^{-1})$ & 86.90$\pm$0.48&6\\
        \hline
    \end{tabular}
    \end{scriptsize}
    \caption{$\mu$TAFS: 10x10 mean cross-validation accuracy (\emph{10x10 CV}) complemented with its standard error for the best model in each data set. The \emph{Classifier} column indicates the best method along with best parameters.}
    \label{Tab:ACCY}
\end{table}

The obtained solutions are displayed in Table \ref{Tab:ACCY}. Among
the eight classifiers used to test the solutions, only the final model
is presented. \emph{Lung Cancer}, \emph{Leukemia} and \emph{Prostate
  Cancer} reach remarkably high accuracies, while \emph{Colon Tumor}
and specially \emph{Breast Cancer} show lower 10x10 CV readings. In
all cases, the subset that delivers this performance is considerable
small, having 7 genes or less (and only 3 genes in the \emph{Leukemia}
data set). Moreover, all Wilcoxon test $p$-values signal significant
differences ($p<0.05$) between the best method and all other methods
in the corresponding data set, except for the lSVM vs. LR in
\emph{Colon Tumor} ($p=0.312$).

\subsection{Discussion of the results}

It is a common practice to compare to similar works in the
literature. Unfortunately, the methodological steps are in general
very different, especially concerning resampling techniques, making an
accurate comparison a delicate undertaking. Nonetheless, such a
comparison is presented in Table \ref{Tab:Compare}. Seven references
which are illustrative of recent work are indicated, including
previous work from the authors. In this table the validation method,
the best classifier and the best reported result are detailed (final
accuracy and number of genes involved).

\begin{table*}[!ht]
  \centering
%  \begin{scriptsize}
  \begin{tabular}{l@{~}c@{~}@{~}c@{~}@{~}c@{~}@{~}c@{~}@{~}c@{~}@{~}c@{~}}
  \hline
  \hline
            &               & Colon       &             & Lung        & Breast      & Prostate\\
  Work      &Validation     & Tumor       &  Leukemia   & Cancer      & Cancer      & Cancer\\
  \hline
  \cite{Hamid}(F) &10x10CV  & 89.36       & 97.89       & 98.84       & 83.37       & 93.43\\
                  &         & (9,3NN)   & (2,NB)    & (4,LR)    & (12,lSVM) & (3,10NN)\\
  \cite{Rui09}(F) &200-B.632& 88.75       & 98.2        & $-$           & $-$           & $-$\\
                  &         & (14,lSVM)& (23,lSVM)& $-$           & $-$           & $-$\\
  \cite{Rui06}(W) &10x10CV  & 85.48       & 93.40       & $-$           & $-$           & $-$\\
                  &         & (3,NB)    & (2,NB)    & $-$           & $-$           & $-$ \\
  \cite{Li08}(W)  &100-RS   & 87.31       & $-$           & 72.20       & $-$           & $-$\\
                  &         & (94,SVM)  & $-$           & (23,SVM)  & $-$           & $-$ \\
  \cite{Bu07}(W)  &50-HO    & 77.00       & 96.00       & 99.00       & 79.00       & 93.00\\
                  &         & (33,rSVM)& (30,rSVM)& (38,rSVM)& (46,rSVM)& (47,rSVM)\\
  \cite{Jin08}(FW)&10x10CV  & $-$           & $-$           & 99.40       & $-$           & 96.30\\
                  &         & $-$           & $-$           & (135,5NN) & $-$           & (79,5NN)\\
  \cite{Rat08}(F) &10CV     & $-$           & 98.6        & 99.45       & 68.04       & 91.18\\
                  &         & $-$           & (2,SVM)     & (5,SVM    ) & (8,SVM)     & (6,SVM)\\
  \hline
  \hline
  \end{tabular}
%  \end{scriptsize}
  \caption{Best results reported in the literature for the explored
    problems: (\textbf{F}) indicates that the referenced work uses a
    Filter-Based Algorithm, (\textbf{W}) for wrapper and
    (\textbf{F-W}) for a combination of both schemes; in parentheses,
    the size of the subset (number of genes) and the inducer
    optimized. A $-$ sign indicates that the problem was not studied
    by the reference. The validations are: 10x10 CV (10 times 10-fold
    Cross Validation), 10 CV (10-fold Cross Validation), 100-RS (100
    times Random Subsampling), 50-HO (50 times Holdout) and 200-B.632
    (0.632 bootstrap of size 200).}
\label{Tab:Compare}
\end{table*}

The \emph{Colon Tumor} data set presents difficulties in
classification, never reaching $90\%$. The solution delivered by
$\mu$-TAFS is comparable with the best known (that of $BGS^3$
\cite{Hamid}); however, it uses 5 genes against the 9 used by
$BGS^3$. The other difficult problem seems to be \textit{Breast
  Cancer}. In this data set, $\mu$-TAFS gives the best result among
the references consulted, using also less genes and in front of
solutions that employ a pure wrapper strategy. For the other three
problems, $\mu$-TAFS is also able to yield better solutions compared
to other approaches, many of them using a much bigger gene subset.

\begin{figure}[!h]
\centering
\begin{tabular}{@{~}c@{~}c@{~}}
    \epsfig{file=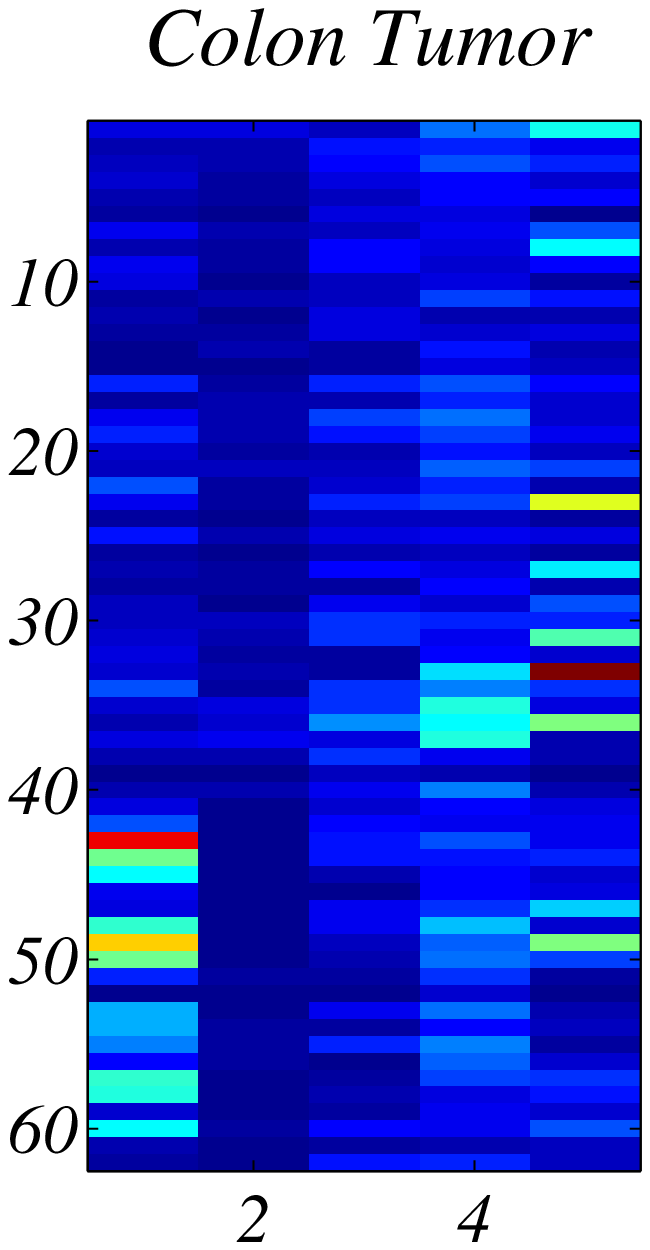,height=3.5cm,bbllx=206,bblly=231,bburx=389,bbury=558}&\epsfig{file=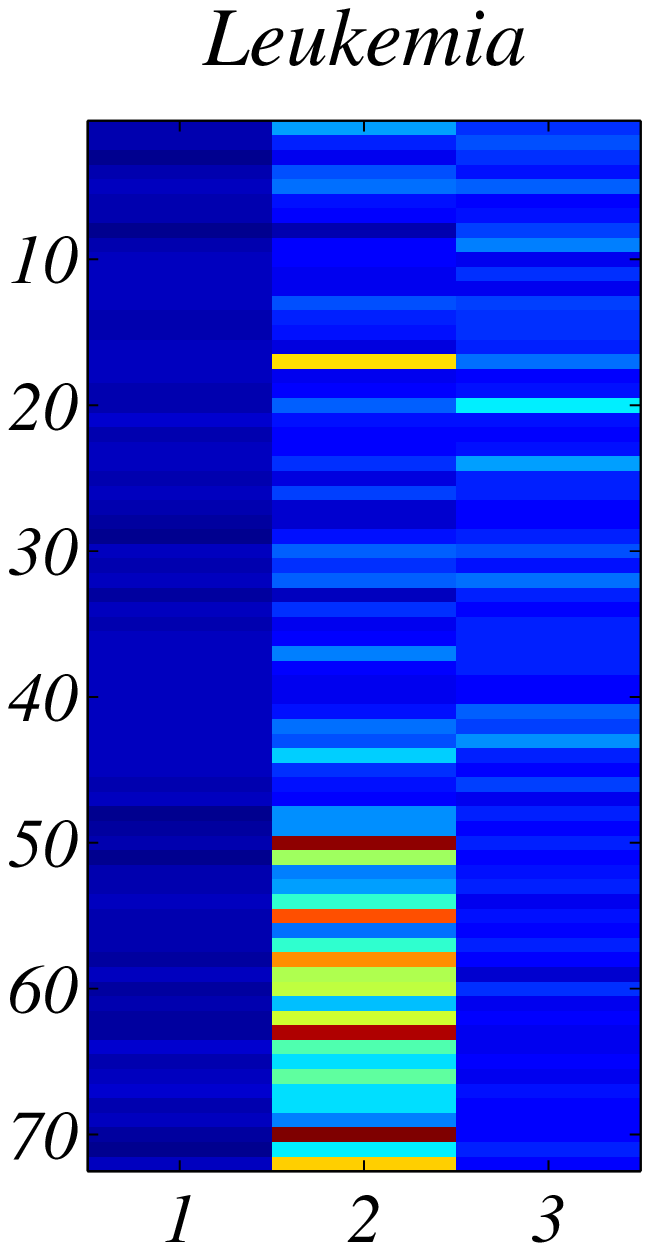,height=3.5cm,bbllx=206,bblly=231,bburx=389,bbury=558}\\
    &\\
    &\\
    \epsfig{file=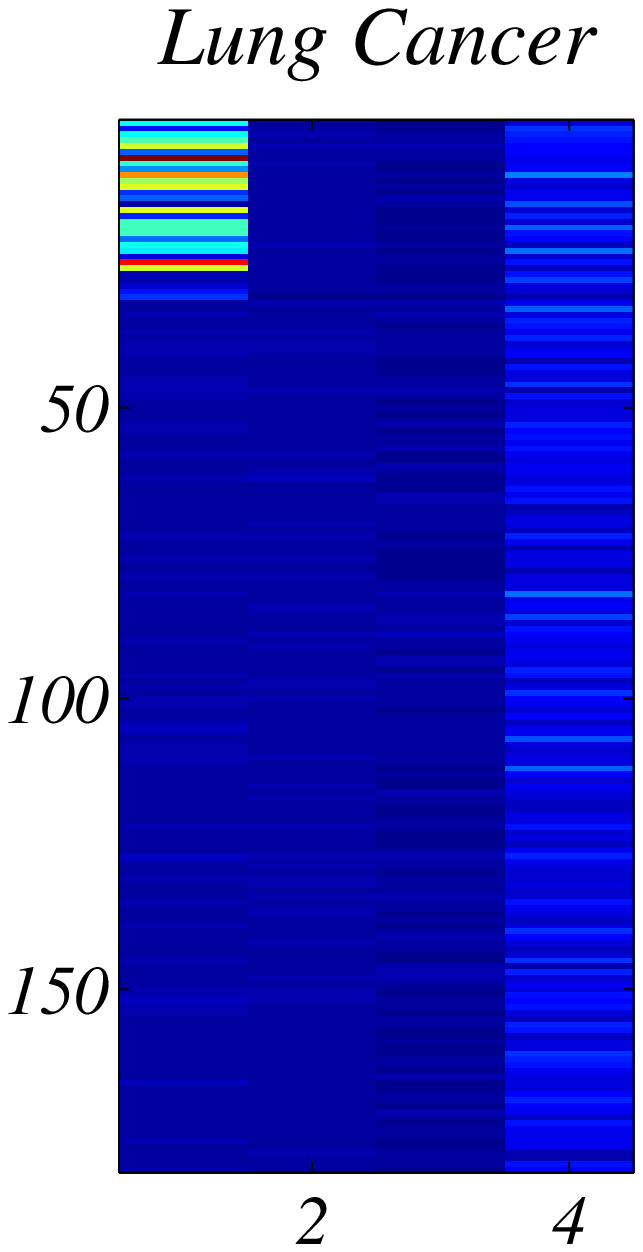,height=3.5cm,bbllx=206,bblly=231,bburx=389,bbury=558}&\epsfig{file=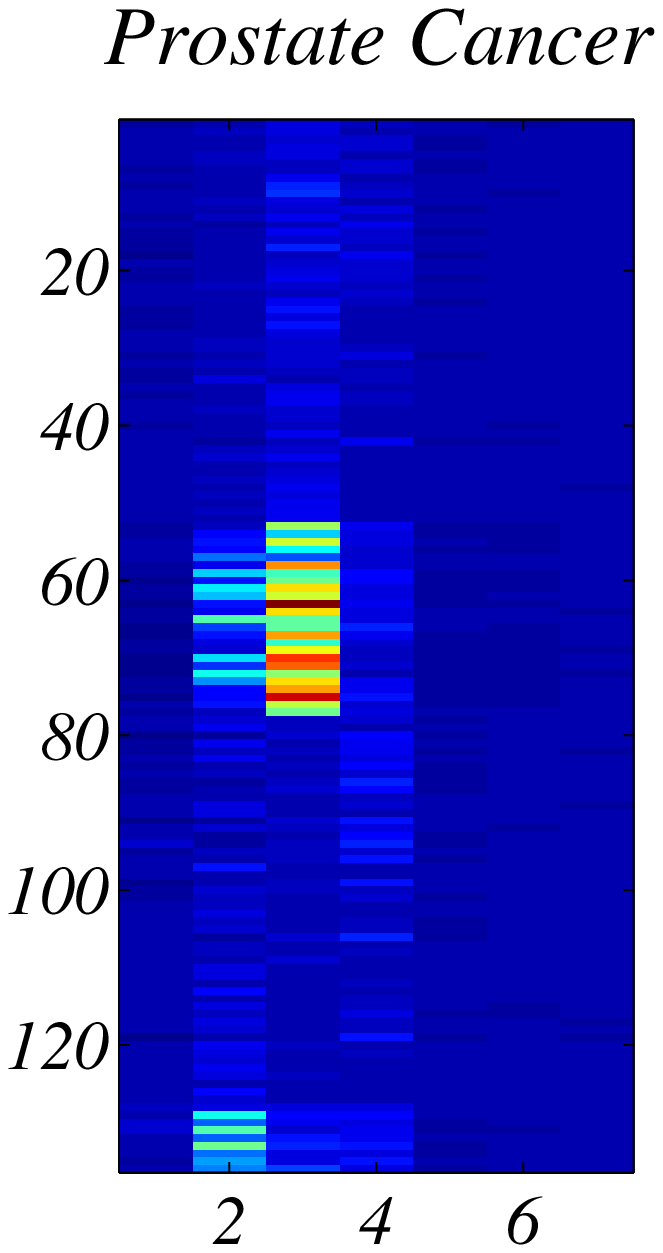,height=3.5cm,bbllx=206,bblly=231,bburx=389,bbury=558}\\
    &\\
    \multicolumn{2}{c}{\epsfig{file=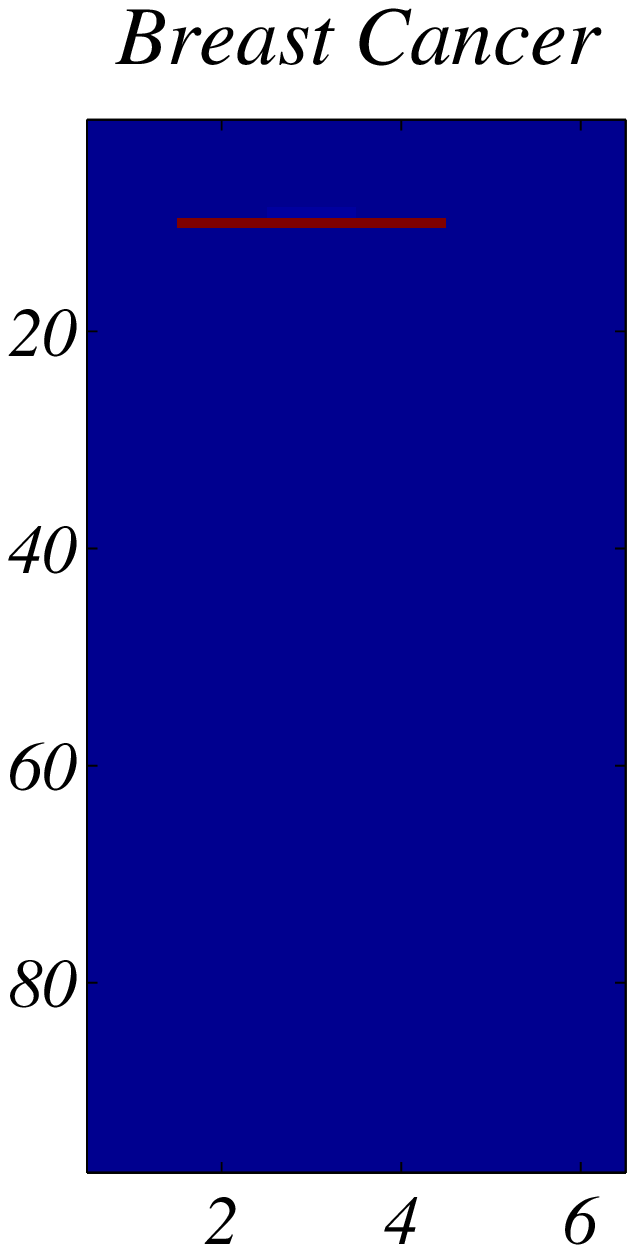,height=3.5cm,bbllx=206,bblly=231,bburx=389,bbury=558}}\\
\end{tabular}
    \caption{Expression levels of winner models as indicated in Table V. Samples for each data set are distributed as follows: \emph{Colon Tumor}: Tumor 1-41, Normal 42-62; \emph{Leukemia}: Tumor 1-48, Normal 49-72; \emph{Lung Cancer}: Tumor 1-31, Normal 32-181; \emph{Prostate Cancer}: Tumor 1-78, Normal 79-136; and \emph{Breast Cancer}: Tumor 1-46, Normal 47-97;}
    \label{FIG:TFS-VIS}
\end{figure}

Expression levels for each model in the five data sets are
given in Fig. \ref{FIG:TFS-VIS}. It is seen that each model posses
genes that are visually identified as the ones that present irregular
expression levels: \emph{Colon Tumor} genes M76378 and T51288;
\emph{Leukemia} genes AFFX-CreX-5\_at and L09209; \emph{Lung Cancer}
gene 37157\_at; \emph{Prostate Cancer} genes 38322\_at and 37639\_at;
and \emph{Breast Cancer} genes Contig14882\_RC, Contig53822\_RC and
Contig57657\_RC.

% en el caption era \ref{Tab:GENES}

\begin{table}[!h]
    \begin{center}
    \begin{tabular}{l l}
        Data set       &   Gene ID\\
        \hline
        \hline
        Colon Tumor     & M76378, H08393, T51849, M19311, T51288 \\
        Leukemia        & AFFX-CreX-5\_at, L09209, X75755\\
        Lung Cancer     & 37157\_at, 33221\_at, 107\_at, 40790\_at\\ %PROBE SETS
        Prostate Cancer & 1230\_g\_at, 38322\_at, 37639\_at, 32909\_at, 660\_at\\
                        &  35998\_at, 34107\_at\\
        Breast Cancer   & AB014543, Contig14882\_RC, Contig53822\_RC\\
                        & Contig57657\_RC, Contig53713\_RC, NM\_006191\\
        \hline
    \end{tabular}
    \end{center}
    \label{Tab:GENES}
    \caption{Genes identification for each final model.}
\end{table}

\subsection{Biological evidence in the solution subsets}
The genes corresponding to the solutions displayed in Table IV are
detailed in Table VI. In the following, known biological evidence is
presented about the effect of gene expressions in each cancer
disease. This evidence is assembled by examining recent relevant
medical literature.

%%%%%%%%%%%%%%%%%%%%%%%%%%%%%%%%%%%%%%%%%%%%%%%%%%%%%%%%%%%
%\noindent
%{\bf Colon Tumor}
%%%%%%%%%%%%%%%%%%%%%%%%%%%%%%%%%%%%%%%%%%%%%%%%%%%%%%%%%%%

\subsubsection*{Colon Tumor}

\begin{itemize}
\item \textbf{M76378} \emph{CSRP1-Cysteine and glycine-rich protein
    1}. This gene encodes a member of the cysteine-rich protein (CSRP)
  family. It may be involved in regulatory processes important for
  development and cellular differentiation. Hypomethylation, a
  decrease in the epigenetic methylation of cytosine and adenosine
  residues in DNA, of CSRIP1 and other genes were confirmed in the
  cancer cells by bisulfite sequencing \cite{Wang08}.

    \item \textbf{H08393} \emph{COL11A2-collagen, type XI, alpha 2
        (Homo sapiens)}. Two alpha chains of type XI collagen, a minor
      fibrillar collagen are encoded by this gene
      \cite{NCBI}. Up-regulation of this gene in the mucosa
      stromal cells of both familial adenomatosis polyposis and
      sporadic colorectal cancer has been detected \cite{Bow08}.

    \item \textbf{T51849} \emph{EPHB1-Tyrosine-protein kinase receptor
        elk precursor}. EphB1 is a member of receptor tyrosine kinases
      of the EphB subfamily and has been positively identified in the
      development, progress and prognosis of colorectal cancers
      \cite{Sheng08}.

    \item \textbf{M19131} \emph{CALM2-calmodulin 2 (phosphorylase
        kinase, delta)}. Caml2 plays an important role in
      intracellular calcium signaling, which regulates a variety of
      cellular processes, such as cell proliferation and gene
      transcription \cite{Bhatta04}. Increased expression levels of
      this gene were found in anaplastic large cell lymphoma cell
      lines \cite{BJH:BJH5816}.

    \item \textbf{T51288} \emph{ASS1-argininosuccinate synthase
        (human)}. Arginine, a semi-essential amino acid in humans, is
      critical for the growth of human cancers as in primary ovarian,
      stomach and colorectal cancer, whose expression levels read high
      values \cite{IJC:IJC25202}.
\end{itemize}

\subsubsection*{Leukemia}
\begin{itemize}
    \item \textbf{AFFX-CREX-5\_AT} NOT IDENTIFIED.

    \item \textbf{L09209} \emph{APLP2-amyloid beta (A4) precursor-like protein 2 (Homo sapiens)}. The function of this gene is not fully understood, but it conjectured that may play a role in the regulation of hemostasis \cite{Genecards}. This gene was reported as over-expressed by other scientific literature as in \cite{17877806}

    \item \textbf{X95735\_at} \emph{ZYX-ZYXIN}. It is involved in the spatial control of actin assembly and in the communication between the adhesive membrane and the cell nucleus \cite{GeneAtlas}. This is a gene found in many cancer classification studies \cite{Go99, Chu05, Sou09}, and is highly correlated with acute myelogenous leukemia.
\end{itemize}

\subsubsection*{Lung Cancer}
\begin{itemize}
    \item \textbf{37957\_at} \emph{ATG4-Autophagy related 4 homolog A}. Autophagy is the process by which endogenous proteins and damaged organelles are destroyed intracellularly. Autophagy is postulated to be essential for cell homeostasis and cell remodeling during differentiation, metamorphosis, non-apoptotic cell death, and aging \cite{Genecards}. It is activated during amino-acid deprivation and has been associated with neurodegenerative diseases, cancer, pathogen infections and myopathies \cite{Rut07}.

    \item \textbf{33221\_at} \emph{PAXIP1-PAX interacting (with transcription-activation domain) protein 1}. Member of the paired box (PAX) gene family, this gene plays a critical role in maintaining genome stability by protecting cells from DNA damage \cite{Genecards, Mun09}. Analysis of pulmonary adenocarcinomas in experiment GDS1650 in \cite{NCBI} records shows over-expression levels of this gene.

    \item \textbf{40790\_at} \emph{BHLHE40-basic helix-loop-helix family, member e40}. This gene encodes a basic helix-loop-helix protein expressed in various tissues, which is may be involved in the control of cell differentiation \cite{NCBI}. Experiments suggest that loss of DEC1 expression is an early event in the development of lung cancer \cite{PATH:PATH1330}

    \item \textbf{107\_at} \emph{RAB40A-member RAS oncogene family}. This gene encodes a member of the Rab40 subfamily of Rab small GTP-binding proteins that contains a C-terminal suppressors of cytokine signaling box \cite{Genecards}. No medical evidence was found in literature about its role in cancer.
\end{itemize}

\subsubsection*{Prostate Cancer}
\begin{itemize}

\item \textbf{1230\_g\_at} \emph{MTMR11-myotubularin related protein
    11}. Experiments on patients with acute lymphoblastic leukemia and
  with Burkitt lymphoma, three putative oncogenes or tumor suppressor
  genes were found, one of them was the MTMR11 \cite{Star07}.

    \item \textbf{38322\_at} \emph{PAGE4-P antigen family, member 4
        (prostate associated)}. This gene is strongly expressed in
      prostate and prostate cancer; and also expressed in other
      tissues as in testis, fallopian tube, uterus, placenta, as well
      as in testicular cancer and uterine cancer \cite{Genecards}.

    \item \textbf{37639\_at} \emph{HPN-Hepsin}. Hepsin is a cell
      surface serine protease and plays an essential role in cell
      growth and maintenance of cell morphology and it is highly
      related with prostate cancer, benign prostatic hyperplasia
      \cite{Genecards}.

    \item \textbf{32909\_at} \emph{AQP5-aquaporin 5}. Acting as a
      water channel protein, Aquaporins are a family of small integral
      membrane proteins linked to other proteins, whose role is the
      generation of saliva, tears and pulmonary secretions
      \cite{Genecards}. Experiments with cases of normal and
      epithelial ovarian tumors tissues suggest an important role of
      this gene in the tumorigenesis of the latter, and a possible
      relationship with the ascites formation of ovarian carcinoma
      \cite{Yang06}.

    \item \textbf{660\_at} \emph{CYP24A1-cytochrome P450, family 24,
        subfamily A, polypeptide 1}. This gene encodes a member of the
      cytochrome P450 superfamily of enzymes. The cytochrome P450
      proteins catalyze many reactions involved in drug metabolism and
      synthesis of cholesterol, steroids and other lipids
      \cite{Genecards}. This gene has been reported as responsible for
      degradation of the active vitamin D metabolite
      1,25-dihydroxyvitamin D3 which is known to be antimitotic in
      prostate cancer cells \cite{Hesso02}.

    \item \textbf{35998\_at} \emph{Hypothetical protein LOC284244 (LOC284244)}. No evidence found.

    \item \textbf{34107\_at}
      \emph{PFKFB2-6-phosphofructo-2-kinase/fructose-2,6-biphosphatase
        2}. The protein encoded by this gene is involved in the
      synthesis and degradation of fructose-2,6-bisphosphate, a
      regulatory molecule that controls glycolysis in eukaryotes
      \cite{Genecards}. It has been suggested that the induction of
      \emph{de novo} lipid synthesis --process that protects cancer
      cells from free radicals and chemotherapeutics-- by androgen
      requires the up-regulation of HK2 and PFKFB2 \cite{Jong11}.

\end{itemize}

\subsubsection*{Breast Cancer}
\begin{itemize}

    \item \textbf{AB014543} \emph{CLUAP1-clusterin associated protein 1 (Homo sapiens)}. This gene is highly expressed in osteosarcoma, ovarian, colon, and lung cancers \cite{Ishi07}.

    \item \textbf{Contig57657\_RC} \emph{PAK1-p21 protein
        (Cdc42/Rac)-activated kinase 1 (Homo sapiens)}. This gene
      encodes a family member of serine/threonine p21-activating
      kinases, known as PAK proteins, whose role is the regulation of
      cell motility and morphology \cite{NCBI}. Pak1 is directly
      related with the Etk/Bmx protein, acting this later as a control
      to the proliferation and tumorigenic growth of mammary
      epithelial cancer cells \cite{Bagheri-Yarmand03082001}.

    \item \textbf{NM\_006191} \emph{PA2G4-Proliferation-associated
        2G4, 38kDa (PA2G4)}. Also known as EBP1, this gene encodes an
      RNA-binding protein that is involved in growth regulation
      \cite{Genecards}. The EBP1 has been shown to be a
      transcriptional corepressor that inhibits the growth of human
      breast cancer cell lines \cite{Akin08}.

    \item \textbf{Contig14882\_RC}, \textbf{Contig53822\_RC}, \textbf{Contig53713\_RC} NOT IDENTIFIED.

\end{itemize}

\section{Conclusions}
A new algorithm for feature selection using Simulated Annealing guided
by the discrete multivariate joint entropy has been introduced and
evaluated. Our experimental results concern the search for small
subsets of highly relevant genes in five public domain Microarray Gene
Expression data samples. The very promising results indicate that
the algorithm offers a promising and general framework for feature
selection in very high dimensional data sets.

The entropic relevance measure has shown to be a good candidate as the
objective function to be optimized by the algorithm. The reported
classification results are competitive to current standards in
analyzing microarray gene expression data with a very affordable
execution time. This last aspect should not be overlooked, since
database size is constantly growing and the complexity of optimization
scenarios (that make extensive use of resampling methods) is ever
greater.

\section{Acknowledgments}
\footnotesize{Authors wish to thank to Spanish CICyT Project
  no. CGL2004-04702-C02-02, CONACyT and UABC for supporting this
  research from its beginning.}

\bibliographystyle{unsrt}
\bibliography{referencias}

\end{document}